\documentstyle[aps,epsfig]{revtex} 
\newcommand{\be}{\begin{equation}}
\newcommand{\ee}{\end{equation}\noindent}
\newcommand{\eei}{\end{equation}}
\newcommand{\bea}{\begin{eqnarray}}
\newcommand{\eea}{\end{eqnarray}\noindent}
\newcommand{\eeai}{\end{eqnarray}}

\def\eq#1{(\ref{#1})}
\begin{document}
\draft
\title{Wave-function renormalization for the Coulomb-gas \\
        in Wegner-Houghton's RG method}
\author{I. N\'andori$^{1,4}$, J. Polonyi$^{2,3}$ and K. Sailer$^4$}
\address{$^1$ Institut f\"ur Theoretische Physik,
Technische Universit\"at Dresden, Germany}
\address{$^2$ Institute of Theoretical Physics, 
Louis Pasteur University, Strasbourg, France}
\address{$^3$ Department of Atomic Physics, 
Lorand E\"otv\"os University, Budapest, Hungary}
\address{$^4$ Department of Theoretical Physics, 
University of Debrecen, Hungary}
\date{\today}
\maketitle
\begin{abstract}
The RG flow for the sine-Gordon model is determined by means of the
method of Wegner and Houghton in next-to-leading order of the
derivative expansion. For small values of the fugacity this agrees
with the well-known RG flow of the two-dimensional Coulomb-gas found in the
dilute gas approximation and a systematic way of obtaining higher-order
corrections to this approximation is given.
\end{abstract}

\section{Introduction}

The  well-known phase structure of the two-dimensional Coulomb-gas was 
investigated in several papers by using the differential RG approach 
with a smooth cut-off in the momentum space (or equivalently using a 
sharp cut-off in the coordinate space)  
(Kosterlitz 1973, Kosterlitz 1974, Jose 1977, Huang 1991, Gersdorff 2000).
Our goal is in this work to determine the RG flow for the two-dimensional 
Coulomb-gas in the framework of the  RG approach of Wegner and Houghton 
(Wegner 1973) taking the field-independent wave-function renormalization 
into account. Results of two different methods were compared: 
renormalization by means of the blocking construction in the 
coordinate space using the dilute gas approximation (real space RG) 
(Kosterlitz 1973, Kosterlitz 1974, Jose 1977, Huang 1991), 
and the renormalization of the equivalent sine-Gordon scalar field model, 
performing the blocking transformations in the momentum space in  
Wegner-Houghton's framework with a sharp cut-off and usage of the 
derivative expansion (Wegner-Houghton RG method) 
(Wegner 1973, Hasenfratz 1986, Polonyi 2000).

It is believed that several different models, like the sine-Gordon, 
Thirring, and the X-Y planar models belong to the same universality
class, namely to that  of  the two-dimensional Coulomb-gas. 
The X-Y model with external field, which is a classical two-component 
spin-model described by the action,
\begin{equation}
S={1\over T} \sum_{x,x'} \cos(\theta_{x}-\theta_{x'}) + 
{h\over T} \sum_{x} \cos(\theta_{x}) 
\end{equation}
has topological excitations, called vortices, which interact via Coulomb 
interaction. Therefore, the X-Y model can be mapped by means of the 
Villain-transformation  to a Coulomb-gas (Huang 1991). Such a mapping 
is, however, only valid up to irrelevant interaction terms. 
The other example, the sine-Gordon model in dimension $d=2$
is a one-component scalar field theory with periodic self-interaction, 
which is defined by the Euclidean action:
\begin{equation}\label{SG}
   S = \int d^2 x \left[{1\over2}(\partial \phi)^2+u\cos(\beta \phi)\right].
\end{equation}
The equivalence between the X-Y model and the lattice regulated 
compactified sine-Gordon model is shown by expressing \eq{SG} 
in terms of the compact variable $z(x)=e^{i\beta\phi(x)}$ (Huang 1991).
This makes the kinetic energy periodic and introduces vortices in the 
dynamics.

There having been made several efforts during the last two decades to
improve the Coulomb-gas results obtained in the dilute-gas approximation, 
where the analogy with the electrodynamics of polarizable media has been 
exploited (Amit 1980, Minnhagen 1985, Kupferman 1997). Here we
also show that the method of Wegner and Houghton  applied to the 
sine-Gordon model provides a systematic way of obtaining higher-order 
corrections to the dilute-gas results.

\section{Real space RG}

The real space RG approach for the Coulomb-gas in dimension $d=2$ 
has been investigated in a great detail in the literature 
(Kosterlitz 1973, Kosterlitz 1974, Jose 1977, Huang 1991).
The dilute vortex gas approximation assumes 
that all the vortex pairs (charges) are separated with at least to 
the minimum distance  $a$ in the coordinate space which is increased 
with infinitesimal steps during the blocking transformation. The 
contributions of the  order higher than $\delta a/a$ are neglected. 
The real space RG equations for the dilute vortex-gas are well-known 
and their derivation is given in the literature
(Kosterlitz 1973, Kosterlitz 1974, Jose 1977, Huang 1991):
\begin{equation}\label{real}
a {d{\tilde h}\over da}  = (2-{T\over 4\pi}) {\tilde h}, \hspace{1cm}
a {dT\over da}  = -\pi T^2 {\tilde h}^2
\end{equation} 
with the  dimensionless coupling constants ${\tilde h}$ and $T$.

\section{Wegner-Houghton RG approach}

The differential RG transformations for the sine-Gordon model are 
realized in the momentum space integrating out the high-frequency  
modes above the moving sharp cut-off $k$ sequentially in infinitesimal 
steps. In the infinitesimal step, corresponding to moving the cut-off 
$k$ to $k-\delta k$, it is integrated for the high-frequency Fourier-modes 
of the field variable $\phi'(x) = \sum_{p=k-\delta k}^{k} \phi_p e^{ipx}$: 
\begin{equation}\label{blocking}
e^{-S_{k-\delta k}[\phi]} = \int D[\phi'] e^{-S_{k}[\phi+\phi']}.
\end{equation}
Performing the path integral in \eq{blocking} using  the saddle 
point expansion at $\phi'$=0, the Wegner-Houghton equation is obtained
(Wegner 1973):
\begin{equation}
\partial_k S_{k}[\phi]
=-{1\over2} tr' \log(G_{p p'}^{-1}[\phi])
\end{equation}
with the inverse propagator 
$G_{pp'}^{-1}[\phi]=\partial^2 S[\phi]/(\partial\phi_p\partial\phi_{p'})$ 
and the trace $tr'$ over the momentum shell $[k-\delta k,k]$. Using the 
derivative expansion in the form 
$S_k=\int d^2 x \left[z(k){1\over2}(\partial \phi)^2+V(\phi,k)\right]$ 
this reduces to:
\begin{equation}\label{momentum}
k\partial_k V(\phi,k)=-k^d \alpha \ln \left({z k^2 + 
{V''(\phi,k)}\over k^2} \right), \hspace{1cm}
k\partial_k z(k)=k^d \alpha [V'''(\phi,k)]^2 
\left[{4 z^2 k^2\over d{\em A}^4} - {z\over{\em A}^3}\right],
\end{equation}
with ${\em A} = (zk^2+V''(\phi,k))$, the potential $V(\phi,k)=u \cos\phi$, 
and the field-independent wave-function renormalization
$z(k)=1/\beta^2$. Eqs. (\ref{momentum}) are obtained by the method described 
in (Polonyi 2000). The flow equations for the dimensionful coupling 
constants $u(k)=-h/T$ and $z(k)=1/T$ are obtained by
expanding both sides of Eqs. (\ref{momentum}) in Fourier series
and neglecting the higher harmonics missed on the  left hand sides.
It is straightforward to use the derivative of the first equation in
\eq{momentum} with respect to $\phi$.
Thus, we find
\begin{equation}\label{dimful}
k\partial_k u(k) = {k^2\over 2\pi} {1\over x(k)} 
        \left[1-\left(1-x^2(k)\right)^{1/2}\right], \hspace{1cm}
k\partial_k z(k) = - {1\over 8\pi} 
        {x^2(k) \left(1+x^2(k)\right) \over \left(1-x^2(k)\right)^{5/2} },
\end{equation}
with $x=u(k)/k^2 z(k)$. In order to compare Eqs. \eq{dimful} with 
Eqs. \eq{real} one should introduce the dimensionless
couplings via $u = -a^{-2} {\tilde h}/T$ and $z=1/T$. Choosing the
moving distance scale according to $1/a = (8\pi^2)^{1/4} k$, we get
\begin{eqnarray}\label{dimless}
a {d{\tilde h}\over da} &=& - {1\over 16\pi^3} {T\over{\tilde h}} 
        \left[1-\left(1-8\pi^2{\tilde h}^2\right)^{1/2}\right] 
        + 2{\tilde h} - \pi T {\tilde h}^3 
        {\left(1+8\pi^2{\tilde h}^2\right) \over 
        \left(1-8\pi^2{\tilde h}^2\right)^{5/2}} , \\ \nonumber
a {dT\over da} &=& - \pi T^2 {\tilde h}^2
        {\left(1+8\pi^2{\tilde h}^2\right) \over 
        \left(1-8\pi^2{\tilde h}^2\right)^{5/2}}.
\end{eqnarray}
Expanding \eq{dimless} in series of ${\tilde h(a)}$, one finds:  
\begin{eqnarray}\label{expand}
a {d{\tilde h}\over da} &=& {\tilde h} 
       \left\lbrack
       2-{T\over 4\pi}
        \left(1+2\pi^2{\tilde h}^2+...\right) - \pi T {\tilde h}^2
        \left(1+...\right)
        \right\rbrack , \\ \nonumber
a {dT\over da} &=& -\pi T^2 {\tilde h}^2 
        \left( 1+28\pi^2{\tilde h}^2+...\right) .
\end{eqnarray}
We see that the leading order terms on the right hand sides are those
of Eqs. (\ref{real}). The terms of higher order in ${\tilde h}$ yield
the infinite series of corrections to the dilute-gas result and are
summed up in a closed form on the right hand sides of Eqs. (\ref{dimless}).

\section{Results}

In a previous paper (Nandori 1999), using the local potential
approximation ($z(k)\equiv 1$), it was shown, that the periodicity and 
the convexity are so strong constraints on the effective
potential $V(\phi,k=0)$ that it becomes flat. This flattening 
was tested numerically, as well. In order to determine the flow 
of the blocked potential in next-to-leading order of the derivative
expansion, the flow of the wave-function renormalization $z(k)$
should also be obtained.

\begin{figure}
  \epsfig{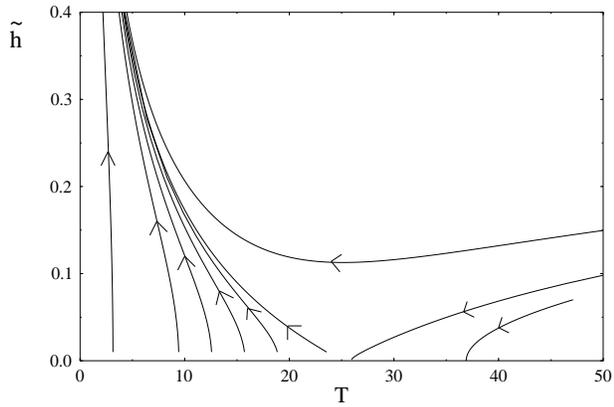}
  \vspace{0.5cm}
  \caption{Phase structure for the two-dimensional Coulomb-gas (or vortex-gas) 
           obtained by the real space RG method using the dilute-gas approximation. 
	   The dimensionless coupling constants are the fugacity of the gas
           (or the external field) $\tilde h$, and the temperature $T$.}
  \label{coulomb}
\end{figure}

The RG flow of $z(k)$ and $u(k)$ (i.e. ${\tilde h}(k)$, $T(k)$), obtained
by the above described two different blocking transformations 
(real space and Wegner-Houghton RG) are qualitatively the same.
The real space RG method (see Eqs.\eq{real}) yielded the well-known 
phase-structure for the Coulomb-gas, (see Fig. \ref{coulomb}) that has already 
been obtained in the literature (Kosterlitz 1973, Kosterlitz 1974, Huang 1991) 
using the dilute gas approximation. There are two phases connected by the 
Kosterlitz-Thouless transition. In the molecular phase  the vortices 
and anti-vortices form closely bound pairs while in the ionised phase 
they dissociate into a plasma. Similar results were obtained in 
(Gersodorff 2000), using momentum space RG with smooth cut-off, without
the dilute gas approximation.

\begin{figure}
  \epsfig{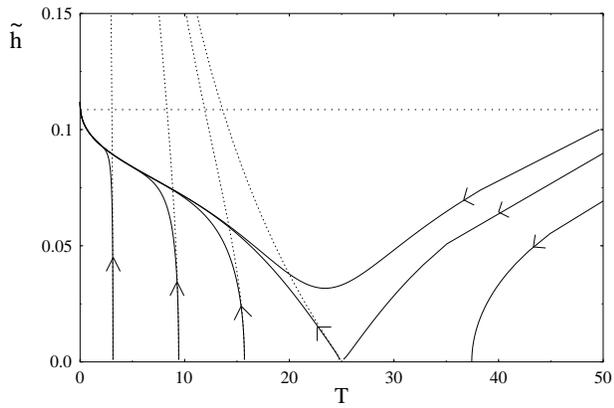}
  \vspace{0.5cm}
  \caption{Phase structure for the sine-Gordon model. The full (dashed) lines 
           correspond to the RG trajectories obtained by the Wegner-Houghton  
           (real space) RG methods. The flow diagram for the sine-Gordon model 
           is valid for $\tilde h_{c}^2  < 8\pi^2$ (see horisontal dotted line),
           above this the Wegner-Houghton equation looses its validity.}
  \label{sinegor}
\end{figure}

The flow diagram for the sine-Gordon model obtained by the 
Wegner-Houghton RG approach (see Eqs.\eq{dimless}) using a sharp 
momentum cut-off, is plotted in Fig. \ref{sinegor}. In order to compare  
the results obtained by  the two different RG methods, four RG trajectories 
calculated by the real space RG method are plotted in
Fig. \ref{sinegor}, as well. For small values of $\tilde h$ and  $T$, 
the trajectories for both methods are the same for the same initial 
conditions. For larger values of $\tilde h$ the dilute gas 
approximation, for large values of $T$ the Villain transformation loose
their validity, so that the different RG trajectories belonging to the same
initial values start to diverge.
The flow diagram for the sine-Gordon model is valid for 
$ \tilde h^2  < 8\pi^2 $. At  $ \tilde h^2  = 8\pi^2 $ a non-trivial 
saddle point occurs in the path integral (\ref{blocking})  and 
the Wegner-Houghton equation looses its validity.

In order to investigate the flattening of the blocked potential
in next-to-leading order of the derivative expansion also the higher
harmonics of the periodic potential should be included. Therefore,
the generalization to the field dependent, periodic wave-function
renormalization should be performed. Since it is unclear how to treat
the field-dependent wave-function renormalization in the
Wegner-Houghton approach, such a generalization is in progress 
using Polchinski's method.

\acknowledgments The authors (I.N. and K.S.) thank G. Soff, G. Plunien 
and R. Sch\"utzhold for the useful discussions. This work has been supported 
by the NATO grant PST.CLG.975722, the DAAD-M\"OB project N$^o$ 27/1999, 
and the grants OTKA T023844/97, T29927/98.

%\begin{references}
%
%
%   \bibitem{kt} J. M. Kosterlitz and D. J. Thouless, J. Phys. {\bf C6} (1973) 118;
%        J. M. Kosterlitz, J. Phys. {\bf C7} (1974) 1046.
%   \bibitem{jkkn} J. V. Jose, L. P. Kadanoff, S. Kirkpatrick, and D. R. Nelson,
%        Phys. Rev. {\bf B16} (1977) 1217.
%   \bibitem{kerson} K. Huang and J. Polonyi, Int. J. Mod. Phys. 
%        {\bf A6} (1991) 409.
%   \bibitem{wet} G.v. Gersdorff and C. Wetterich, {\em Nonperturbative 
%        Renormalization Flow and Essential Scaling \\ 
%        for the Kosterlitz-Thouless Transition}, hep/th-0008114 
%   \bibitem{wh} F.J. Wegner and A. Houghton, Phys. Rev. {\bf A8} (1973) 401;
%   \bibitem{hh} A. Hasenfratz and P. Hasenfratz, Nucl. Phys. {\bf B270}
%          (1986) 687.
%   \bibitem{composite} J. Polonyi and K. Sailer, {\em  Renormalization of
%        composite operators}, hep-th/0011083
%   \bibitem{amit} D.J. Amit, Y.Y. Goldschmidt, and G. Grinstein,
%         J. Phys. {\bf A13} (1980) 585.
%   \bibitem{minn}
%         P. Minnhagen, Phys. Rev. {\bf B32} (1985) 3088.
%   \bibitem{kupf}
%         R. Kupferman and A.J. Chorin, {\em A numerical study of the
%           Kosterlitz-Thouless transition in a two-dimensional 
%         Coulomb or vortex gas}, LBNL-40368/1997.
%   \bibitem{per} I. N\'andori, J. Polonyi, K. Sailer, hep-th/9910167, 
%        accepted to Phys. Rev. D.
%
%
%\end{references}

\section*{REFERENCES}
\noindent
Amit, D.J., Y.Y. Goldschmidt, and G. Grinstein, J. Phys. {\bf A13} 585. (1980)\\
Gersdorff G.v. and C. Wetterich, {\em Nonperturbative Renormalization Flow and 
	Essential Scaling for the Kosterlitz-Thouless Transition}, hep/th-0008114\\
Hasenfratz A. and P. Hasenfratz, Nucl. Phys. {\bf B270} 687. (1986)\\
Huang K. and J. Polonyi, Int. J. Mod. Phys. {\bf A6} 409. (1991)\\
Jose J. V., L. P. Kadanoff, S. Kirkpatrick, and D. R. Nelson, Phys. Rev. {\bf B16} 1217. (1977)\\
Kosterlitz J. M. and D. J. Thouless, J. Phys. {\bf C6} 118. (1973)\\
Kosterlitz J. M., J. Phys. {\bf C7} 1046. (1974)\\
Kupferman R. and A.J. Chorin, {\em A numerical study of the Kosterlitz-Thouless 
	transition in a two-dimensional Coulomb or vortex gas}, LBNL-40368/1997.\\
Minnhagen P., Phys. Rev. {\bf B32} 3088. (1985)\\
N\'andori I., J. Polonyi, K. Sailer, hep-th/9910167, accepted to Phys. Rev. D\\
Polonyi J. and K. Sailer, {\em  Renormalization of composite operators}, hep-th/0011083\\
Wegner F.J. and A. Houghton, Phys. Rev. {\bf A8} 401. (1973)\\

\section*{FIGURE CAPTIONS}

\end{document}